\documentstyle[prd,aps]{revtex}
\begin{document}
\title{
 Physical Implications of a Vector-like Extension of the Standard Model\footnote{Talk presented at  KOSEF-JSPS Winter School on Recent 
Developments in Particle and Nuclear Theory, Seoul, Korea, February 21-28, 
1996 (To be published in Supplement of Journal of Korean Physical Society) }}

\author{ Kazuo Fujikawa}
\address{
Department of Physics, University of Tokyo\\
         Bunkyo-ku, Tokyo  113, Japan}

\maketitle
\begin{abstract}
Physical implications of a vector-like extension of the standard model 
for heavier quarks and leptons with $SU(2)\times U(1)$ gauge symmetry and 
only one Higgs doublet are discussed. This scheme 
incorporates infinitely many
fermions, and the chiral structure is realized by a non-vanishing analytic 
index of mass matrices. The model is perturbatively well controllable, and
 the physical Higgs particle generally mediates leptonic as well as quark 
flavor-changing processes at a rate below the present experimental limit.
This model illustrates a general feature of models where some of the fermion 
masses are of  non-Higgs origin.
\end{abstract}

\section{Introduction}

\par
  An abrupt end of  proliferation of quarks
and leptons(in particular, light neutrinos) at the 3rd generation
is rather mysterious.  On the other hand, the weak coupling of the top quark 
is known 
to be almost purely left-handed on the basis of the $\rho$-parameter
analysis[1] and also by an analysis of the $b\rightarrow s\gamma$ decay[2].
One of the reasons why the generations with heavier  masses
are prohibited may be the dynamical stability of  Weinberg-Salam
theory with chiral (left-handed) weak couplings. Some time ago,
we performed an analysis of this problem associated with heavy fermions[3]. 
We here discuss this problem  on the  basis of a concrete model[4].

In the standard model, all the masses of
gauge bosons and fermions are generated by the Higgs mechanism.
In the following, we examine whether
heavier quarks and leptons with masses of non-Higgs origin in  TeV region, 
for example, can be accommodated in the standard model when perturbed by the 
Higgs mechanism which generates masses for conventional quarks and leptons. 
  
  We first recall  a consequence of the coupling of  W boson
to the fermion doublet $ \psi_{k},k=1,2$ generically defined by
\begin{equation}
{\cal
L}=(1/2)g\bar{\psi}_{k}(T^{a})_{kl}\gamma^{\mu}(a+b\gamma_{5})\psi_{l}W_{\mu}^{a}
\end{equation}
The longitudinal coupling of $W_{\mu}^{a}$, which is related to Higgs
mechanism , is studied by replacing $W_{\mu}^{a}\rightarrow (2/gv)\partial
_{\mu}S^{a}(x)$ in (1) with $ S^{a}(x)$ the unphysical Higgs scalar.
We then obtain
\begin{equation}
{\cal L}=a(\frac{m_{k}-m_{l}}v)\overline{\psi}_{k}(T^{a})_{kl}\psi_{l}S^{a}
+b(\frac{m_{k}+m_{l}}v)\overline{\psi}_{k}(T^{a})_{kl}\gamma_{5}\psi_{l}S^{a}
\end{equation}
by using the equations of motion.

The typical mass scale of the Higgs world is
\begin{equation}
v=247GeV
\end{equation}
and thus
\begin{equation}
   b(m_{k}+m_{l})\gg v  ,\ or ,\   a|m_{k}-m_{l}|\gg v \\
\end{equation}
induces a strongly interactiong sector into the standard model.
We here regard the situation in (4) as unnatural.In other words, the standard 
model with chiral couplings does not accommodate  fermions with masses much
larger than the Higgs scale (3).
The heavier fermions in the standard scheme, if they should exist without
introducing strong couplings,  should thus
have almost pure vector-like couplings (i.e., $b \simeq 0 $ in (4))
and that the mass of the fermion doublet should be almost degenerate
(i.e., $a|m_{k}-m_{l}| \leq v$ in (4)).
If  heavier quarks and leptons satisfy the above conditions, they
have no sizable couplings to the Higgs sector. In other words, their
masses primarily come from dynamics
which is different from  the Higgs mechanism in the standard model.
Also, the breaking mechanism of SU(2) (both of local as well as custodial)
in the standard model
is concluded to be a typical phenomenon in the energy scale of $v$.

\section{A Vector-like Extension of the Standard Model}
\par
We consider an $SU(2){ \times} U(1)$ gauge theory
written in an abbreviated notation[4]
\begin{equation}
{\cal L}_{L}=\overline{\psi}i\gamma^{\mu}D_{\mu}\psi
            - \overline{\psi}_{R}M\psi_{L}
            - \overline{\psi}_{L}M^{\dagger}\psi_{R}
\end{equation}
with
\begin{equation}
\not{\!\! D}=\gamma^{\mu}(\partial_{\mu} - igT^{a}W_{\mu}^{a}
            - i(1/2)g^{\prime}Y_{L}B_{\mu})
\end{equation}
and  $Y_{L}=1/3$ for quarks and $Y_{L}=-1$ for leptons. The field
$\psi$ in (5) is a column vector consisting of an infinite number of
$SU(2)$ doublets, and
the infinite dimensional $nonhermitian$ mass matrix $M$ satisfies the
index condition
\begin{equation}
\dim\ker\ M = 3,\ \dim\ker\ M^{\dagger}=0
\end{equation}
In the  explicit "diagonalized" expression of $M$
\begin{eqnarray}
M&=&\left(\begin{array}{ccccccc}
          0&0&0&m_{1}&0    &0    &..\\
          0&0&0&0    &m_{2}&0    &..\\
          0&0&0&0    &0    &m_{3}&..\\
          .&.&.&.    &.    &.    &..
          \end{array}\right)
\end{eqnarray}
the fermion $\psi$ is written as
\begin{equation}
 \psi_{L}=(1-\gamma_{5})/2\left(
 \begin{array}{c}
  \psi_{1}\\ \psi_{2}\\ \psi_{3}\\ \psi_{4}\\.
 \end{array}
 \right), \ \
 \psi_{R}=(1+\gamma_{5})/2\left(
 \begin{array}{c}
  \psi_{4}\\ \psi_{5}\\ \psi_{6}\\.\\.
 \end{array}
 \right)
\end{equation}
We thus have 3 massless left-handed $SU(2)$ doublets $\psi_{1},\psi_{2},
\psi_{3}$, and an
infinite series of vector-like massive $SU(2)$ doublets $\psi_{4},
\psi_{5},...$ with
masses $m_{1},m_{2},..$ as is seen in

\begin{eqnarray}
{\cal L}_{L}&=&\bar{\psi}_{1}i\not{\!\! D}(\frac{1-\gamma_{5}}{2})\psi_{1}
               +\bar{\psi}_{2}i\not{\!\! D}(\frac{1-\gamma_{5}}{2})\psi_{2}
                \nonumber\\
            & &+\bar{\psi}_{3}i\not{\!\! D}(\frac{1-\gamma_{5}}{2})\psi_{3}
                \nonumber\\
            & &+\bar{\psi}_{4}(i\not{\!\! D} -m_{1})\psi_{4}
               +\bar{\psi}_{5}(i\not{\!\! D} -m_{2})\psi_{5} + ...
\end {eqnarray}

An infinite number of right-handed fermions in a doublet notation are also
introduced by( again in an abbreviated notation)
\begin{equation}
{\cal L}_{R}=\overline{\phi}i\gamma^{\mu}(\partial_{\mu}-i(1/2)g^{\prime}
Y_{R}B_{\mu})\phi - \overline{\phi}_{L}M^{\prime}\phi_{R}
-\overline{\phi}_{R}(M^{\prime})^{\dagger}\phi_{L}
\end{equation}
where
\begin{equation}
Y_{R}=\left(\begin{array}{cc}
            4/3&0\\
            0&-2/3
            \end{array}\right)
\end{equation}
for quarks and
\begin{equation}
Y_{R}=\left(\begin{array}{cc}
            0&0\\
            0&-2
            \end{array}\right)
\end{equation}
for leptons, and the mass matrix $M^{\prime}$   satisfies the index
condition
(7) but in general it may have different mass eigenvalues from
those in(8). After the diagonalization of $M^{\prime}$, $\phi$ is
written as
\begin{equation}
 \phi_{L}=(1-\gamma_{5})/2\left(
 \begin{array}{c}
  \phi_{4}\\ \phi_{5}\\ \phi_{6}\\ .\\ .
 \end{array}
 \right), \ \
 \phi_{R}=(1+\gamma_{5})/2\left(
 \begin{array}{c}
  \phi_{1}\\ \phi_{2}\\ \phi_{3}\\ \phi_{4}\\ .
 \end{array}
 \right)
\end{equation}
Here, $\phi_{1}, \phi_{2}$,and  $ \phi_{3}$ are right-handed and massless,
and $\phi_{4}, \phi_{5},....$ have masses $m_{1}^{\prime}, m_{2}^{\prime}$,..
\begin{eqnarray}
{\cal L}_{R}&=&\bar{\phi}_{1}i\not{\!\! D}(\frac{1+\gamma_{5}}{2})\phi_{1}
               +\bar{\phi}_{2}i\not{\!\! D}(\frac{1+\gamma_{5}}{2})\phi_{2}
                \nonumber\\
            & &+\bar{\phi}_{3}i\not{\!\! D}(\frac{1+\gamma_{5}}{2})\phi_{3}
                \nonumber\\
            & &+\bar{\phi}_{4}(i\not{\!\! D} -m_{1}^{\prime})\phi_{4}
               +\bar{\phi}_{5}(i\not{\!\! D} -m_{2}^{\prime})\phi_{5} + ...
\end {eqnarray}
with
\begin{equation}
   \not{\!\! D}= \gamma^{\mu}(\partial_{\mu}-i(1/2)g^{\prime}
                 Y_{R}B_{\mu})
\end{equation}

The present model is vector-like and manifestly anomaly-free
before the  breakdown  of parity (7);after the breakdown of
parity,the model still stays anomaly-free provided that both of $M$ and
$M^{\prime}$ satisfy the index condition (7).  Unlike  conventional vector-like
models with a finite number of components, the present scheme avoids
the
appearance of a strongly interacting right-handed sector despite of the
presence of heavy fermions.
A truncation of the present scheme to a finite number of heavy
fermions (for example, to only one heavy doublet in $\psi$) is still
consistent, although it is no more called vector-like.

The massless fermion sector in the above scheme  reproduces the same
 set of fermions
as in the standard model. However, heavier fermions have distinct
features. For example, the heavier fermion doublets with the smallest
masses are
described by
\begin{eqnarray}
{\cal L}&=&\overline{\psi}_{4}i\gamma^{\mu}(\partial_{\mu}-igT^{a}W_{\mu}^{a}
           -i(1/2)g^{\prime}Y_{L}B_{\mu})\psi_{4}-m_{1}\overline{\psi}_{4}
           \psi_{4}\nonumber\\
        & &+\overline{\phi}_{4}i\gamma_{\mu}(\partial_{\mu}
           -i(1/2)g^{\prime}Y_{R}B_{\mu})\phi_{4}
           -m_{1}^{\prime}\overline{\phi}_{4}\phi_{4}
\end{eqnarray}
The spectrum of fermions is thus $doubled$ to be vector-like in the
sector  of heavy fermions and ,at the same time, the masses of $\psi$ and
$\phi$ become non-degenerate, i.e., $m_{1}{\neq}m_{1}^{\prime}$ in general.
As a result, the fermion number anomaly
is generated only
by the first 3 generations of light fermions;the violation
of baryon number is not enhanced by the presence of heavier fermions.
The masses  of  heavy doublet components in $\psi$  are degenerate
 in the present zeroth order approximation. The masses of heavy doublets in
$\phi$ are also taken to be degenerate for simplicity.

In the present scheme we distinguish two classes of chiral symmetry
breaking;one which is
related to the breaking of gauge symmetry (Higgs mechanism), and the
other which is related to the mass of heavier fermions but not related to
the breaking of gauge symmetry. The transition from one class of
chiral symmetry breaking to the other, which is also accompanied by the
transition from chiral to vector-like gauge couplings, is  assumed
to take place at the mass scale of the order of $v$ in (3). In any case
if the $SU(2){\times}U(1)$ gauge symmetry should be universally valid
regardless of the magnitude of the mass of fermions, just like
electromagnetism and gravity, the coupling of heavier fermions is
required to become vector-like:\ Heavy gauge bosons can naturally
couple to light fermions, but the other way around imposes a
stringent constraint on the chirality of fermions.
We are  here interested
in the possible on-set of heavier fermions at the order of a few TeV,
although these vector-like components are often assumed to acquire
masses of the order of grand unification scale(Georgi's survival
hypothesis[7]).

\section{ Higgs Perturbation and  Light Fermion Masses }

\par
As for the mass generation of the first 3 generations of quarks and
leptons and also the custodial $SU(2)$ breaking of heavier fermions,
one  may introduce a Yukawa interaction for quarks, for example, in
an abbreviated notation
\begin{eqnarray}
{\cal L}_{Y}&=&\bar{\psi}_{L}G_{u}\varphi\phi_{R}^{(u)}
         + \bar{\psi}_{L}G_{d}\varphi^{c}\phi_{R}^{(d)}\nonumber\\
            &+&\bar{\psi}_{R}G_{u}^{\prime}\varphi\phi_{L}^{(u)}
         + \bar{\psi}_{R}G_{d}^{\prime}\varphi^{c}\phi_{L}^{(d)} + h.c.
\end{eqnarray}
where  $\varphi(x)$ is the conventional Higgs doublet ( and
$\varphi(x)^{c}$ is its conjugate) , and
$G_{u}, G_{d}, G_{u}^{\prime}$ and $G_{d}^{\prime}$ are
infinite dimensional
coupling matrices acting on $\psi$ and $\phi$. Corresponding to the
presence of only one $W$-boson, we here assume the existence of only
one Higgs doublet.
The fields $\bar{\psi}_{L}$ or $\bar{\psi}_{R}$ in (18) stands for the
doublets in (5), and $\phi_{R}^{(u)}$ (or $\phi_{L}^{(u)}$) and
$\phi_{R}^{(d)}$ (or $\phi_{L}^{(d)}$), respectively, stand for the
upper and lower components of the doublets $\phi$ in (11). If one
retains only the first two terms and their conjugates in (18) and if
only the massless components of $\psi_{L}$ and $\phi_{R}$ in (9)
and (14) are considered, (18) reduces to the Higgs coupling of the
standard model.

We  postulate that the coupling matrices G  ( which generically include
$G^{\prime}$ hereafter)
are such that
the interaction (18) is perturbatively well controllable,namely, the
typical element of coupling matrices $G$ is bounded by the gauge
coupling $g$,
\begin{equation}
    |G|{\leq}g
\end{equation}
By this
way the masses of the first 3 generations of light fermions are generated
from (18) below the mass scale in (3).
For the heavier fermions,
the interaction (18) introduces the breaking of custodial $SU(2)$ and
also  fermion mixing. After the conventional $SU(2)$ breaking,
\begin{equation}
{\langle}\varphi{\rangle}=v/\sqrt{2}
\end{equation}
one may diagonalize the mass matrix
in (18) together with the mass matrices in (5) and (11).This introduces
a generalization of the ordinary fermion mixing matrix .
If one assumes a generic situation,
\begin{eqnarray}
   & & m_{i}, m_{j}^{\prime} {\gg} gv, \ \ \ |m_{i}-m_{j}^{\prime}| {\gg}
        gv{\nonumber} \\
   & & |m_{i}-m_{j}| {\gg} gv, \ \ \ |m_{i}^{\prime}-m_{j}^{\prime}| {\gg}
        gv
\end{eqnarray}
for any combination of (renormalized) heavy fermion masses $m_{i}$ and
$m_{j}^{\prime}$,
the masses of heavier fermions are little modified by the Higgs
coupling.
As a fiducial value of the on-set of heavy fermion mass, we here choose
\begin{equation}
m_{i} {\sim} a\ few\ TeV
\end{equation}

In practical calculations, it is convenient to diagonalize the light
fermion masses in addition to (8) and its analogue of $M^{\prime}$
but leave the mixing of light and
heavy fermions
non-diagonalized, instead of diagonalizing all the masses. In this case
the effects of  heavy fermions on the processes of light fermions are
estimated in the power expansion of $G$. The mass term after diagonalizing
the light quarks in the up-quark sector, for example, is given by
\begin{eqnarray}
(\bar{\psi}_{L},\bar{\Psi}_{L},\bar{\Phi}_{L}){\cal M}\left(
\begin{array}{c}
\phi_{R}\\ \Psi_{R}\\ \Phi_{R}
\end{array}
\right)
+ h.c.
\end{eqnarray}
where the mass matrix ${\cal M}$ has a structure
\begin{eqnarray}
&& \left(\begin{array}{ccccccccc}
m_{u}&0     &0    &     & & &                   &&\\
0    &m_{c}&0     &     &\Large{0}& &&\tilde{G}v/\sqrt{2}&\\
0    &0     &m_{t}&     & & &                   &&\\
     &     &               &m_{1}&0    &0&                   &&\\
     &\tilde{G}v/\sqrt{2}& &0    &m_{2}&0             &    &Gv/\sqrt{2}&\\
     &                  &  &0    &0    &..            &    &            &\\
     &                  & & &          &  &m_{1}^{\prime}&0            &0\\
     &\Large{0}         & & &Gv/\sqrt{2}& &0             &m_{2}^{\prime}&0\\
     &                  & & &           & &0             &0          &..
\end{array}\right)\nonumber
\end{eqnarray}
To avoid introducing further notational conventions,  we here use the
fields $\psi_{L}$ and $\phi_{R}$ in(23) for the first 3 light
(i.e.,massless in the zeroth order approximation) fermion components of
$\psi$ and $\phi$ in (9) and (14),respectively; $\Psi$ and $\Phi$ stand
for the
remaining heavy quark components of $\psi$ and $\phi$ in (9) and (14).
The coupling matrix $\tilde{G}$ is different from the original $G$ by
the unitary
transformation of light quarks performed in the process of diagonalizing
light quark masses. But the order of magnitude of $\tilde{G}$ is still
the same as that of $G$. We note that the Higgs coupling (18) mixes the 
left- and right-handed fermions and thus leads to a breaking of the 
index condition (7), which ensured the presence of 3 massless fermions
without the Higgs coupling.

The physical Higgs $H(x)$ coupling in the unitary gauge is given by the
replacement $ v \rightarrow v + H(x)$
in the above mass matrix(23).

If one sets $G = 0$ and $\tilde{G} = 0$ in the mass matrix(23), the
light and heavy quark sectors become completely disconnected, not only
in the Higgs coupling but also in the gauge coupling except for the
renormalization effects due to heavy quark loop diagrams. This means
that the \underline{direct} effects of heavy fermions on the processes
involving light quarks and leptons only can be calculated as a power
series in $\tilde{G}$ and $G$, provided that these effects of heavy
fermions are small. It is shown [4] that these effects are in fact of
controllable magnitude if the condition $|G| \leq g$ in (19) is satisfied
and the mass spectrum of heavy fermions starts at a few TeV.

Some of the physical implications of the present scheme have 
benn analyzed [4][5].
 Heavier fermions
are expected to decay mainly into the Higgs particle and light fermions
with natural decay width
\begin{equation}
 \Gamma_{i}{\sim}|G|^{2}m_{i}.
\end{equation}
if we take the Higgs mass at the "natural" value
$ m_{H}\sim v $ with $v$ in (3).

The mass spectrum of light fermions is influenced by heavier fermions
through the mixing in (18), but  the most natural choice $|G| \sim g$ already
gives a sensible result [4],although a certain fine tuning is
required to
account for the actual masses of the electron and up and down quarks.

A characteristic feature of the present extension of the standard
model  is that the leptonic as well as quark
flavor is generally violated; this breaking is caused by the mixing of
light and heavy fermions in (18). The diagonalization of mass matrix
does not diagonalize the Higgs coupling in general unlike the standard
model, and the physical Higgs particle at the tree level
also mediates flavor changing processes although its contribution is not
necessarily a dominant one; this is a rather general feature of a model 
where some of the fermion masses are of non-Higgs origin.
In the limit of large heavy fermion masses
$m_{i} \rightarrow \infty$ in the present model, this flavor changing coupling vanishes.
The Higgs and heavy fermion contributions to GIM suppressed processes
 in the standard model are shown to be small below the present experimental
limit[4].

The leptonic flavor changing processes such as $K^{0}_{L} \rightarrow
e\bar{\mu}$ are also induced by the mixing of heavy fermions ,
as was noted above.
The decay rate of $K_{L}^{0}\rightarrow e\bar{\mu}$ is then estimated at the
order
\begin{equation}
\Gamma(K_{L}^{0}\rightarrow e\bar{\mu})\leq 10^{-8}\times
\Gamma(K_{L}^{0}\rightarrow \mu\bar{\mu})
\end{equation}
where $\Gamma(K_{L}^{0}\rightarrow \mu\bar{\mu})$ is given by the
standard model.

It is confirmed that
$CP$ violation does not appear in the zeroth order
approximation without the Higgs coupling (see eq.(10))
and it arises solely in the Higgs sector(18);the pattern of
CP violation becomes more involved than in the standard model and
CP phase is no more limited to W couplings.

As for the neutrinos, the first three neutrinos will remain
massless if one assumes the absence of
right-handed components(i.e., if $\phi$ in (11) is a singlet).
We however expect the appearance
of vector-like heavy neutrinos above TeV region.  

\section{Discussion}
\par
A vector-like extension of the standard model examined in this note is
natural, in the sense that the validity of perturbation theory (19) combined 
with  a sensible choice of heavy fermion mass scale (23) lead to consistent 
results as a first order approximation.

The present model as it stands is, however, a  phenomenological
one:The appearance of many fermions with vector-like couplings might
be natural from some kind of composite picture of fermions, or if the
fermions are elementary their masses might  arise from a topological
origin as is suggested by (7) or from some kind of space-time
compactification. But the fundamental issue of the   breaking mechanism of
chiral and parity symmetries by the analytic index  in (7) needs to be 
explained.
The breaking of asymptotic freedom of QCD by heavy fermions becomes
appreciable only at the mass scale of these heavy fermions due to
the decoupling phenomenon.


\begin{thebibliography}{1}
\bibitem{1}
M. Veltman, Nucl. Phys. {\bf B123},89(1977).\\
R.D.Peccei, S.Peris and X.Zhang, Nucl. Phys. {\bf B349},305(1991).

\bibitem{2}
K. Fujikawa and A. Yamada, Phys. Rev. {\bf D49},5890(1994), and references 
therein.

\bibitem{3}
K. Fujikawa,Prog. Theor. Phys. {\bf 61},1186(1979).

\bibitem{4}
K. Fujikawa, Prog. Theor. Phys. {\bf 92},1149(1994).\\
For somewhat related models of fermions from different view points,see\\
J. Pati,Phys. Rev. {\bf D51}, 2451(1995). \\
K. Inoue, Prog. Theor. Phys. {\bf 93}, 403(1995).\\
H. Zheng, Phys. Rev. {\bf D52},6500(1995); hep-ph/9602340.


\bibitem{5}
We here assume that
the mass spectrum of heavier fermions is rather sparsely distributed.
We thus estimate the effects of the lightest heavier fermions on
physical processes involving ordinary fermions in the standard model.

\bibitem{6}
G. 't Hooft, Phys. Rev. Lett. {\bf 37},8(1976).

\bibitem{7}
H. Georgi, Nucl.Phys.{\bf B156},126(1979).

\end{thebibliography}
\end{document}